\documentclass{article}
\usepackage{float}
\usepackage{float}
\usepackage{float}
\usepackage{arxiv}

\usepackage[utf8]{inputenc} 
\usepackage[T1]{fontenc}    
\usepackage{hyperref}       
\usepackage{url}            
\usepackage{booktabs}       
\usepackage{amsfonts}       
\usepackage{nicefrac}       
\usepackage{amsthm}         
\theoremstyle{remark}
\newtheorem{remark}{Remark}

\usepackage{amsmath}        
\usepackage{microtype}      
\usepackage{amsmath}        
\usepackage{amssymb}        
\usepackage{appendix}      
\usepackage{amsthm}         
\usepackage{amsthm}         
\usepackage{amsthm}         
\usepackage{amsthm}
\newtheorem{theorem}{Theorem}
\newtheorem{proposition}{Proposition}
\newtheorem{statement}{Statement}
\DeclareMathOperator*{\argmax}{arg\,max}

\usepackage{amsthm}         
\usepackage{lipsum}
\usepackage{graphicx}
\graphicspath{ {./images/} }

\title{Incorporating Taxonomies of Cyber Incidents into Detection Networks for Improved Detection Performance.}

\author{
 Ryan Warnick \\
  Microsoft Security Research\\
  Microsoft\\
  Redmond, W.A. 98052 \\
  \texttt{ryanwarnick@microsoft.com} \\
}

\usepackage{float}
\begin{document}
\maketitle
\begin{abstract}
Many taxonomies exist to organize cybercrime incidents into ontological categories. We examine some of the taxonomies introduced in the literature; providing a framework, and analysis, of how best to leverage different taxonomy structures to optimize performance of detections targeting various types of threat-actor behaviors under the umbrella of precision and recall. Networks of detections are studied, and results are outlined showing properties of networks of interconnected detections. Some illustrations are provided to show how the construction of sets of detections to prevent broader types of attacks is limited by trade-offs in precision and recall under constraints. An equilibrium result is proven and validated on simulations, illustrating the existence of an optimal detection design strategy in this framework.
\end{abstract}


\section{Introduction}\label{sec1}

Cybercrime is a multi-billion dollar a year criminal enterprise, with attacks occurring an estimated 2244 \cite{attacksperday} times on every device, every day, as far back as 2007; with these numbers likely greatly increasing in the intermediate time window. The World Bank cites that cybercrime incident direct costs are expected to reach 251 billion USD globally by 2031 \cite{world_bank}; with direct costs entailing tangible financial losses, damages, and hardships endured by victims. Additionally, estimates are provided for the expected individual cost of a data breach to increase to 4.35 million USD per breach. As a consequence, the cybersecurity industry has grown at an exponential rate year over year, with cybercrime remediation being a large part of this sector.

Cybercrime remediation is contingent on accurate detection of criminal activity, and detection work is a significant point of investment for cybersecurity vendors; with time to remediation being of utmost importance \cite{review_time_factor} \cite{real_time_remediation}. Taxonomies for cybersecurity incidents have achieved increasing interest in recent years as the difficulty of navigating the complexity of various computing and networking architectures makes pinning down a uniform ontological mapping between a cyberattack and a categorization difficult. However, having a taxonomy allows one to more appropriately and efficiently tailor remediation and response strategies \cite{taxonomy_of_taxonomies}. Ways to categorize taxonomies can be found in \cite{taxonomy_of_taxonomies}; with a thorough review of older taxonomies, including links to github repositories for presented taxonomy architectures, available in \cite{taxonomy_review}. These taxonomies and ontologies provide a schematic for placing cybersecurity incidents in general populations and sub-populations, and allow the design of effective detections targeting specific categories of attack strategies. The work presented here seeks to assess networks of detections targeting different locations on the taxonomy in terms of their individual, as well as joint, performance.

Various measures of performance exist to quantify how well a binary detection is doing. These are well summarized in statistical literature. They include Signal to Noise Ratio (SNR) \cite{SNR}, Precision and Recall \cite{precision_recall}, the F1 Score (F1) \cite{Rijsbergen1979}, and many others \cite{performance_metrics}. The most common metric by far to measure the accuracy of detections in cybersecurity settings is SNR, as this can be readily retrieved from feedback from individual Security Operations Centers (SOCs). However, this metric suffers as a poor descriptor of the performance of a detection, due to not properly assessing accuracy and retrieval ranges in a multifaceted way.

For this work, we examine precision and recall as counter-balanced performance metrics, in the context of cyberthreat taxonomy architectures such as the MITRE ATT\&CK\cite{mitre_attack} framework. The goal of the work is to provide a common set of tools to understand how a taxonomy and set of logical detections built on said taxonomy can be optimized for semi-local (a local set of detections) and global (the complete set of detections) performance. Additionally, results are proven showing the existence of a Nash equilibrium \cite{nash1950} between detections in a detection set acting as noncooperative agents trying to maximize the squared sum of precision and recall in a noncooperative game; leading to an interpretation of optimal detection performance.

The remainder of the article proceeds as follows: Section \ref{sec2} introduces some preliminary notation and concepts to understand the rest of the article; with Section \ref{subsec1} introducing notation related to the taxonomy and detections, Section \ref{subsec2} introducing the notation related to the performance metrics, and Section \ref{subsec3} introducing the concept of a conditional detection and notation for that aspect of the work. Section \ref{sec3} contains the principal results of the work. Section \ref{sec4}. Section \ref{sec:discussion} describes the principal contributions of the article; with possible future directions outlined at the end of that section. An Appendix (Section \ref{appendix}) contains the proofs of the theorems.

\section{Notation and Basic Concepts}\label{sec2}
We introduce some notation here to assist in navigating the article. Section \ref{subsec:taxonomy} introduces the framework for modeling the taxonomical structure and detections, and relevant notation. Section \ref{subsec:performance} illustrates the notation and concepts for the relevant performance metrics. Section \ref{subsec:conditioning} outlines dependency structures between detections common in SOC operations. Section \ref{subsec:graphs} ties this all together with the overlaid network of detections, and consolidates the notation of the detections into a directed acyclic graph.

\subsection{Notation for Taxonomy and Detections}\label{subsec1}
\label{subsec:taxonomy}
\begin{figure}[h]
\centering
\includegraphics[width=0.35\textwidth]{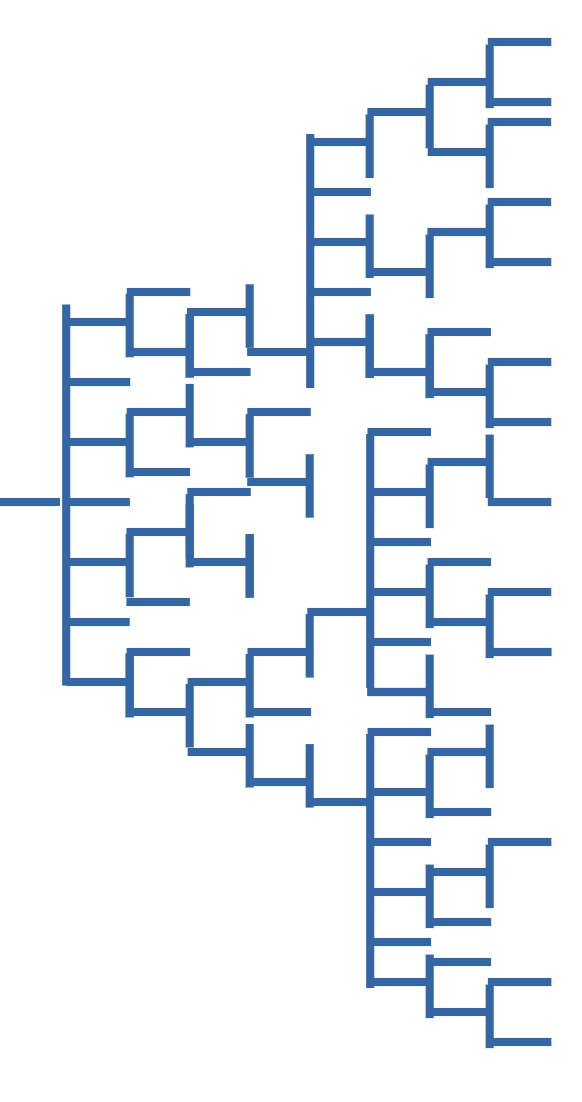}
\caption{Illustration of a taxonomy tree for a given attack strategy and potential leaves (endpoints) of detections in the taxonomy. The taxonomy has ten levels with each branch branching off into new leaves or branches.}\label{fig1}
\end{figure}

A diagram of a simple taxonomy for an event is provided in Figure \ref{fig1}, with a starting level in the taxonomy leading into several potential branches or leaves, with a leaf terminating and a branch leading to other potential branches or leaves.

Assume each branch is an ordered set. Our current location in the taxonomy is denoted by $T_{\epsilon_1}$ where $\epsilon_1$ denotes the current branch we are on. We then partition the current branch into associated leaves or branches by appending the index of the next leaf or branch $\epsilon_2$, to navigate to $T_{\epsilon_1\epsilon_2}$, where $\epsilon_2$ is a branch or leaf of $\epsilon_1$. This is repeated until an endpoint is reached: $T_{\epsilon_1,\dots,\epsilon_n}$. Note that $n$ is always less than or equal to the depth of the taxonomy.

Each branch or leaf represents an opportunity for a detection, with detections at increasingly low levels in the hierarchy. Indicate a detection by $D_{\epsilon_1,\dots_{\epsilon_n}}$.

\subsection{Notation for Performance Metrics}\label{subsec2}
\label{subsec:performance}
Given a detection $D_{\epsilon_1,\dots,\epsilon_n}$ for arbitrary $n\leq$ the depth of the taxonomy, we denote by $FP$ the total number of false positives of the detection, $FN$ the total number of false negatives of the detection, $TP$ the total number of true positives of the detection, and $TN$ the total number of true negatives of the detection.

Precision and Recall are statistical measurements indicating different types of performance of a particular detection \cite{precision_recall}. In layman's terms, precision is how targeted a detection is, and recall is how broad of a stroke the detection takes in what it signals on. The formulae for these two quantities are as follows:

\begin{equation}
Precision = \frac{TP}{TP + FP}\label{precision}
\end{equation}

\begin{equation}
Recall = \frac{TP}{TP + FN}\label{recall}
\end{equation}

\begin{figure}[H]
\centering
\includegraphics[width=0.6\textwidth]{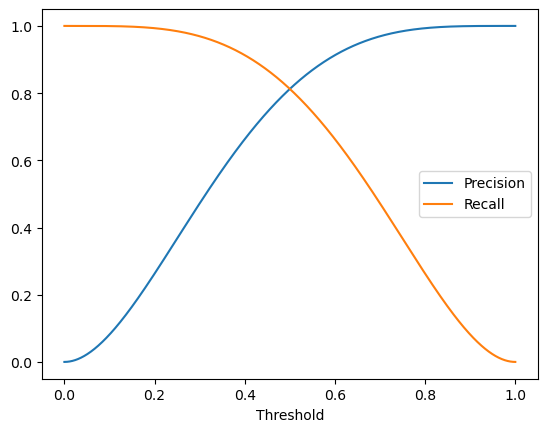}
\caption{Illustration of reciprocal relationship between precision and recall of statistical detection with threshold between 0 and 1.}\label{fig2a}
\end{figure}

\begin{figure}[H]
\centering
\includegraphics[width=0.6\textwidth]{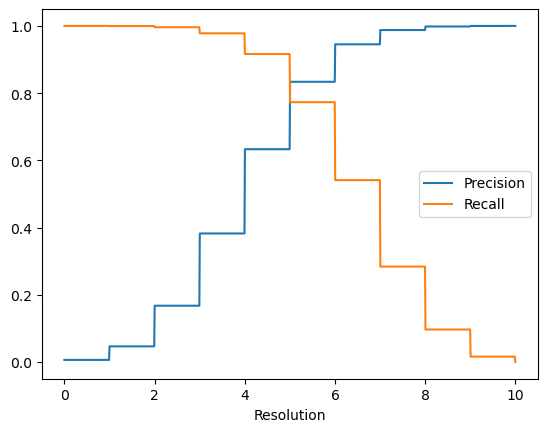}
\caption{Illustration of reciprocal relationship between precision and recall for logical filter of increasing resolution.}\label{fig2b}
\end{figure}

Figure \ref{fig2a} and \ref{fig2b} give illustrations, for toy binary detections, of the reciprocal relationship between Precision and Recall. Figure \ref{fig2a} shows a detection with a quantitative threshold between 0 and 1, with higher thresholds indicating more parsimonious selection of when to fire. As precision goes up, recall goes down; and vice versa. Figure \ref{fig2b} instead shows a logical detection (if X occurs fire the signal), and as the resolution of the filter (the specificity of its steps in the logical chain) goes up, precision goes up; and conversely recall goes down. 

A thorough overview of performance metrics for binary classifiers can be found in \cite{performance_metrics}; including but not limited to Precision and Recall.

We denote the precision and recall of a detection by applying the operators $Precision(\circ)$ and $Recall(\circ)$ in a conceptual scenario to the detection $D_{\epsilon_1,\dots,\epsilon_n}$.

It is important to note that precision and recall can be thought of as estimates from a conceptual scenario of conditional probabilities \cite{conditional_probabilities}. Let $Y_{\epsilon_1,\dots,\epsilon_n} \in \{P,N\}$ denote the true class of the signal at $T_{\epsilon_1,\dots,\epsilon_n}$, and $D_{\epsilon_1,\dots,\epsilon_n}\in\{P,N\}$ denote the estimated class. Then:

\begin{align}\label{precision_conditional_probability}
Precision(D_{\epsilon_1,\dots,\epsilon_n}) &= P(Y_{\epsilon_1,\dots,\epsilon_n}= P | D_{\epsilon_1,\dots,\epsilon_n} = P)\\
Recall(D_{\epsilon_1,\dots,\epsilon_n}) &= P( D_{\epsilon_1,\dots,\epsilon_n} = P|Y_{\epsilon_1,\dots,\epsilon_n}= P)\label{recall_conditional_probability}
\end{align}

We work in the rest of this paper with idealized operators for precision and recall, $\lim_{N\rightarrow \infty} Precision(D_{\epsilon_1,\dots,\epsilon_n})$ and $\lim_{N\rightarrow \infty} Recall (D_{\epsilon_1,\dots,\epsilon_n})$; equating to what would happen under the true underlying distribution $P(Y_{\epsilon_1,\dots,\epsilon_n})$ and $P(D_{\epsilon_1,\dots,\epsilon_n})$ in Equations \ref{precision_conditional_probability} and \ref{recall_conditional_probability}. Or, in other words:

\begin{equation}
Precision(D_{\epsilon_1,\dots,\epsilon_n})(\lambda) = \lim_{N\rightarrow \infty} \frac{TP(\lambda)}{TP(\lambda)+FP(\lambda)}
\label{limit_precision}
\end{equation}

and

\begin{equation}
Recall(D_{\epsilon_1,\dots,\epsilon_n})(\lambda) = \lim_{N\rightarrow \infty} \frac{TP(\lambda)}{TP(\lambda)+FN(\lambda)}
\label{limit_recall}
\end{equation}

For a particular threshold of the detection $\lambda$.

Note that the precision and recall of a detection depend upon what threshold we use (either quantitative or logical). Some detections will have multiple thresholds that have to be set (having some logical and some quantitative statements, or multiple expressions all of the same type). 

Denote by $\Lambda_{\mathcal{D}}=\prod_{D\in\mathcal{D}}\Lambda_D$ the cross product of the set of all possible values for each individual threshold that needs to be set for detection $\mathcal{D}$, and $\lambda_{D}$ compact interval contained individual element of $\Lambda_{D_{\epsilon_1,\dots,\epsilon_n}}$. We assume that $\Lambda_D$ is continuous (or upper hemi-continuous) for all $D\in\mathcal{D}$:

\begin{equation}
Precision(D)(\lambda_D) = \frac{P(Y_D= P, D= P)(\lambda_D)}{P(D = P)(\lambda_D)}
\label{eq::precision_threshold}
\end{equation}

and

\begin{equation}
Recall(D)(\lambda_D) = \frac{P(Y_D= P, D= P)(\lambda_D)}{P(Y_D = P)}
\label{eq::recall_threshold}
\end{equation}

are piecewise continuous or piecewise constant over $\lambda_D$.

\subsection{Conditioned Detections}\label{subsec3}
\label{subsec:conditioning}
A detection can be influenced by another detection. That is, if a detection has a logical filter in it based on whether another detection has fired or not. Denoting a pair of detections $D^{(1)}_{\epsilon_1,\dots,\epsilon_n}$ and $D^{(2)}_{\epsilon_1,\dots,\epsilon_m}$, we denote that $D^{(2)}_{\epsilon_1,\dots,\epsilon_m}$ is conditioned on $D^{(1)}_{\epsilon_1,\dots,\epsilon_n}$ by writing, in an abuse of probability notation, $D^{(2)}_{\epsilon_1,\dots,\epsilon_m} |  D^{(1)}_{\epsilon_1,\dots,\epsilon_n}$. We are using the operator $|$ to denote a \textit{design constraint} (operational conditioning). Not a stochastic conditional distribution.In the conditional detection scenario, optimizing a detection to have improved Precision or Recall influences the Precision or Recall of dependent detections. From here forward we assume all relevant detections are contained in the set $\mathcal{D}$

This equates to a change in measure of $D^{(2)}_{\epsilon_1,\dots,\epsilon_m} |  D^{(1)}_{\epsilon_1,\dots,\epsilon_n}$ on the probability distribution over $D^{(2)}_{\epsilon_1,\dots,\epsilon_m} |  D^{(1)}_{\epsilon_1,\dots,\epsilon_n} =P$ vs. $D^{(2)}_{\epsilon_1,\dots,\epsilon_m} |  D^{(1)}_{\epsilon_1,\dots,\epsilon_n} =N$, where the mass in the binary random variable (dependent upon $\lambda$) is renormalized in $D^{(2)}_{\epsilon_1,\dots,\epsilon_m} |  D^{(1)}_{\epsilon_1,\dots,\epsilon_n}$ conditional on $D^{(1)}_{\epsilon_1,\dots,\epsilon_n}$, similar to a typical binary random variable/binary random variable probabilistic relationship.

\begin{figure}[H]
\centering
\includegraphics[width=0.99\textwidth]{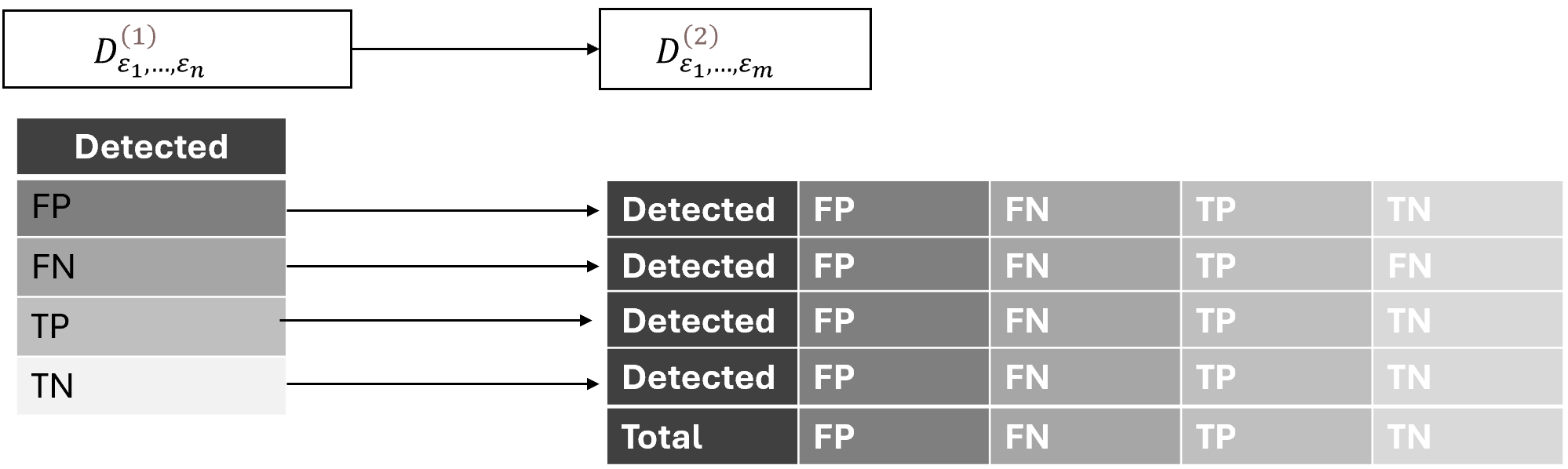}
\caption{An illustration of how a logical detection can be conditioned on another logical detection, and how the event of a FN, TN, FP, or TP in $D^{(1)}_{\epsilon_1,\dots,\epsilon_n}$ and influence the outcome in $D^{(2)}_{\epsilon_1,\dots,\epsilon_m}$.}\label{fig3}
\end{figure}

Figure \ref{fig3} provides an illustration of what conditional detections signify. The dependencies between the $FN$, $TN$, $FP$, and $TP$ of $D^{(2)}_{\epsilon_1,\dots,\epsilon_m}$ are dependent on what occurs with $D^{(1)}_{\epsilon_1,\dots,\epsilon_n}$.

Additionally, detections can depend upon sets of other detections. This constructs a type of structure analogous to a Bayes Network \cite{Pearl1988} of conditional detections. Where a detection $D$ is conditioned on another set of detections, and we denote its conditioning set $\mathcal{P}(D)$ and term it the parent set. This gives us the following notational representation:

\begin{equation}
D | \mathcal{P}(D) \label{eq:conditional_network}
\end{equation}

This creates, as opposed to the $4\times 4$ table in Figure \ref{fig3}, the cross product of $m+1$ times 4 dimensional vectors to create a table of dimension $m+1$. Each axis has 4 elements; with each element being an $m+1$-tuple of $FP$, $FN$, $TN$, and $TP$.

We introduce a concept here to make the proofs in Section \ref{sec3} more appropriately general in concordance with the framework just discussed.

\paragraph{Operational gating.}
A detector $D$ fires operationally (\cite{thomas1973boolean}) if its internal rule $R_D(\lambda_D)$ is positive \emph{and} its gate condition over parents holds. We analyze two canonical gates:
\[
\textsc{AND:}\quad D^{\mathrm{op}}_{\lambda}= R_D(\lambda_D) \wedge \bigwedge_{C\in\mathcal{P}(D)} C^{\mathrm{op}}_{\lambda_C};\qquad
\textsc{NOT:}\quad D^{\mathrm{op}}_{\lambda}= R_D(\lambda_D) \wedge \bigwedge_{C\in\mathcal{P}(D)} \neg C^{\mathrm{op}}_{\lambda_C}.
\]
Results extend to mixed monotone gates by isotonicity (Remark on isotonicity and what we mean by internal rule outlined in Remark~\ref{rem:isotone}).

\subsection{Graphs on Detections}
\label{subsec:graphs}
We construct a graph of conditional detections and assume the graph is directed and acyclic. Denote the graph $\mathcal{G} = (\mathcal{D},\mathcal{E})$, where $\mathcal{D}$ is defined as previously, and $\mathcal{E}$ is the edge set of directed edges between the taxonomy detections and other conditioning detections existing as ordered pairs. E.g. $(D_i,D_j) \in \mathcal{E}$, for detections $D_i,D_j \in \mathcal{D}$, if and only if $D_i \in \mathcal{P}(D_j)$.

\begin{figure}[H]
\centering
\includegraphics[width=0.75\textwidth]{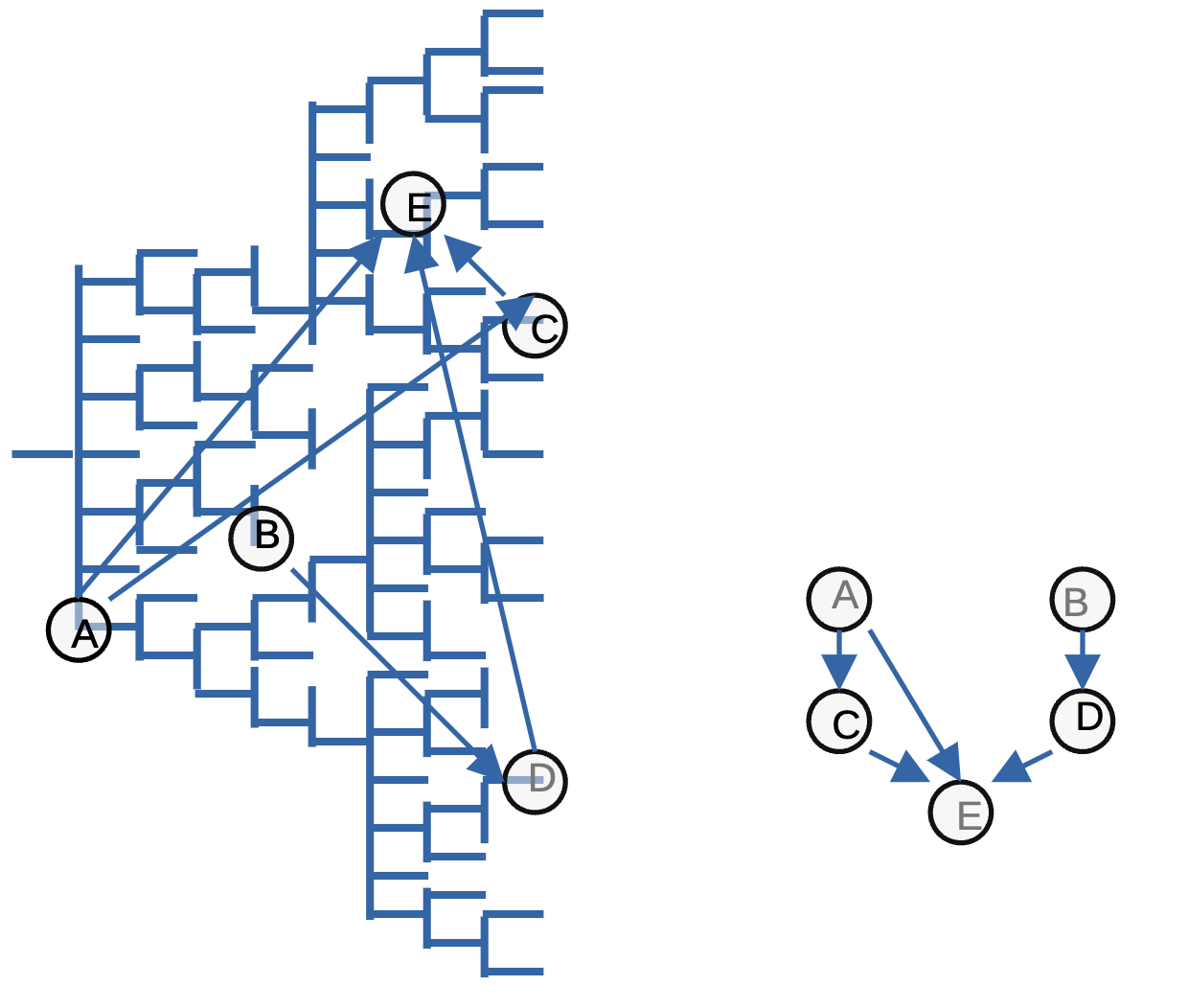}
\caption{Illustration of the graph $G$ generated by the set of conditional detections in the taxonomy presented in Figure \ref{fig1}. The taxonomy to the left illustrates the locations of the detections $A,B,C,D,E$ at$\{T_A,T_B,T_C,T_D,T_E\}$ in the taxonomy, with associated conditional relationships between them illustrated in the graph on the right.}\label{fig5}
\end{figure}

An example of a simple set of detections $A,B,C,D,E$ on the taxonomy presented in Figure \ref{fig1} is provided in Figure \ref{fig5}. Denote These detections have locations in the taxonomy $T_A,T_B,T_C,T_D,T_E$, but the graph existing between them exists as the set $\mathcal{D}=\{A,B,C,D,E\}$ and edge set $\mathcal{E} = \{(A,C),(A,E),(C,E),(B,D),(D,E)\}$. The conditional detection graph is denoted by $\mathcal{G} = (\mathcal{D},\mathcal{E})$.

A complete account of graphical architecture for probabilistic networks is available in \cite{koller2009probabilistic}. However, we note here that our goal is optimal design of detections and not understanding the architecture of the Bayes network. Also of note is that our graph is between detections having varying levels of performance (Precision, Recall, FP, TP, FN, and TN), and this is our principal interest, and not probability distributions with associated marginals and conditionals.
\newline 
\begin{remark}[Isotonicity of gates and Detection Rule]
\label{rem:isotone}
Let $E(\lambda)$ be any gate event that is monotone in each parent (increasing for \textsc{AND}, decreasing for \textsc{NOT}). If the parent’s precision increases while recall does not increase, then the induced change in $E(\lambda)$ preserves the stated monotonic direction for the child’s precision/recall. Mixed gates with monotone Boolean formulas inherit these properties coordinatewise \cite{angeli2003monotone}. The internal rule of detection $D$, denoted by $R_D(\lambda_D)$, is simply saying that the detection's rule to fire (the threshold has been reached) has been activated; whether the parents set's arrangement is in concordance with the gating or not. This allows the the statement outlined above to follow as a logical operation; not a probabilistic one. \cite{thomas1973boolean}
\end{remark}

\section{Results}\label{sec3}

\paragraph{Assumptions.}
Let $\Lambda_D \subset \mathbb{R}$ be a nonempty, compact interval of admissible thresholds for detector $D$. 
For each fixed profile of other detectors' thresholds $\lambda_{-D}$, define the population
\[
\mathrm{Prec}_D(\lambda_D\,;\,\lambda_{-D}) := \frac{\mathbb{P}(D_{\lambda}=P \mid Y=P)}{\mathbb{P}(D_{\lambda}=P)}, 
\quad
\mathrm{Rec}_D(\lambda_D\,;\,\lambda_{-D}) := \mathbb{P}(D_{\lambda}=P \mid Y=P),
\]
whenever the denominator is positive.

\textbf{(A1)} For every $D$, $\lambda_D \mapsto \mathrm{Rec}_D(\lambda_D\,;\,\lambda_{-D})$ is continuous on $\Lambda_D$ and $\lambda_D \mapsto \mathrm{Prec}_D(\lambda_D\,;\,\lambda_{-D})$ is upper semicontinuous on $\Lambda_D$. (This allows piecewise-constant or stepwise detectors.)

\textbf{(A2)} For every $D$, there exists $\eta_D>0$ such that $\mathbb{P}(D_{\lambda}=P) \ge \eta_D$ for all $\lambda_D \in \Lambda_D$ and all $\lambda_{-D}$ considered, or else precision is defined on $\{\mathbb{P}(D_{\lambda}=P)>0\}$ and optimized over this closed subset.

\textbf{(A3)} For any conditioning set $\mathcal{P}(D)$ of parents used operationally (gates defined below), $\mathbb{P}(\bigwedge_{C\in\mathcal{P}(D)} Y_{\lambda_C}=P) \ge \eta_{\mathcal{P}}>0$ uniformly on the relevant action sets (or precision is evaluated conditionally on this event, with the same upper semicontinuity).

\begin{theorem}
[\textbf{Continuity/semi-continuity under gating}]
\label{theorempartialcontinuity}
Under (A1)–(A3) and fixed $\lambda_{-D}$, the map $\lambda_D \mapsto \mathrm{Rec}_{D^{\mathrm{op}}}(\lambda_D\,;\,\lambda_{-D})$ is continuous on $\Lambda_D$. 
Moreover, $\lambda_D \mapsto \mathrm{Prec}_{D^{\mathrm{op}}}(\lambda_D\,;\,\lambda_{-D})$ is upper semicontinuous on $\Lambda_D$, and continuous wherever $\mathbb{P}(D^{\mathrm{op}}_{\lambda}=P)$ stays bounded away from $0$.
\end{theorem}
\begin{proof}
Proof is available in appendix section \ref{proofpartialcontinuity}
\end{proof}
\begin{theorem}
[\textbf{Monotonicity of precision/recall under \textsc{AND} gating}]
\label{thm:mono-and}
Fix a detector $D$ with parents $\mathcal{P}(D)$ under \textsc{AND} gating. Suppose that for every parent $C\in\mathcal{P}(D)$ we move $\lambda_C$ to weakly increase $\mathrm{Prec}_C$ (with recall not increasing). Then for fixed $\lambda_D$, $\mathrm{Prec}_{D^{\mathrm{op}}}$ weakly increases and $\mathrm{Rec}_{D^{\mathrm{op}}}$ weakly decreases. 
\end{theorem}

\begin{proposition}
[\textbf{Monotonicity under \textsc{NOT} gating}]
\label{prop:mono-not}
\label{propositionconditioning}
Under \textsc{NOT} gating, if $\mathrm{Prec}_C$ weakly increases (with recall not increasing) for each parent $C$, then $\mathrm{Prec}_{D^{\mathrm{op}}}$ weakly \emph{decreases} and $\mathrm{Rec}_{D^{\mathrm{op}}}$ weakly \emph{increases}.
\end{proposition}

\begin{proof}
Proofs available in appendix Section \ref{proofmono}.
\end{proof}

Note that the reciprocal is true for Recall as Precision and Recall have a monotonically counter-balanced relationship. To express this rigorously we include it as a proposition and prove it in the Appendix.

\begin{theorem}
[\textbf{Deterministic taxonomy idealization}]
\label{thm:hier-det}
Assume a lossless taxonomy such that every child label implies its ancestor: $Y_n=P \Rightarrow Y_m=P$ for all ancestors $m \prec n$ (no noise). Then for any child $Y_n$ gated by parent $Y_m$ via \textsc{AND},
\[
\mathrm{Prec}_{C_n^{\mathrm{op}}} \ge \mathrm{Prec}_{D_m}, \qquad 
\mathrm{Rec}_{C_n^{\mathrm{op}}} \le \mathrm{Rec}_{D_m}.
\]
\end{theorem}

\begin{theorem}
[\textbf{Stochastic taxonomy, robustness}]
\label{thm:hier-stoch}
If $\mathbb{P}(Y_m=P \mid C_n=P) \ge 1-\epsilon$ for all thresholds in the action sets, then
\[
\mathrm{Prec}_{C_n^{\mathrm{op}}} \ge \mathrm{Prec}_{D_m} - O(\epsilon), \qquad 
\mathrm{Rec}_{C_n^{\mathrm{op}}} \le \mathrm{Rec}_{D_m} + O(\epsilon).
\]
\end{theorem}

\begin{proof}
Proof available in appendix Section \ref{proofroot}.
\end{proof}

\begin{theorem}
[\textbf{Existence of pure-strategy equilibrium}]
\label{thm:existence}
Let $\{\Lambda_D\}_{D=1}^N$ be nonempty, compact, convex intervals. For each $D$, define the utility
\[
U_D(\lambda_D;\lambda_{-D}) = a_D \,\mathrm{Prec}_{D^{\mathrm{op}}}(\lambda_D;\lambda_{-D}) 
+ b_D \,\mathrm{Rec}_{D^{\mathrm{op}}}(\lambda_D;\lambda_{-D}),
\quad a_D,b_D \ge 0,\; (a_D,b_D)\neq (0,0).
\]
Assume (A1)–(A3) and that for every fixed $\lambda_{-D}$, $U_D(\cdot;\lambda_{-D})$ is quasi-concave on $\Lambda_D$ and $U_D$ is continuous on $\prod_D \Lambda_D$.

Then this framework is a noncooperative game with an equilibrium point \cite{debreu1952social,glicksberg1952further,fan1952minimax} where appropriate thresholds can be selected where no detection can be improved without other detections wanting to change their thresholds for improvement. 
\end{theorem}

\begin{proof}
Proof available in appendix Section \ref{proofequilibria}.
\end{proof}

We also want to make a small note here that in scenarios where we're not working with precision and recall, but instead SNR, then if we are trying to maximize SNR, and the idealized version of SNR as a function of the threshold is quasi-concave, then the equilibrium point exists as well.

\section{Simulation Study}\label{sec4}

\subsection{Design of Experiments}\label{subsec4}
\subsection{Simulation Framework}

The simulation experiments are designed to illustrate and validate the theoretical properties of \emph{taxonomy-conditioned detection networks} introduced in the paper---particularly the continuity, monotonicity, and equilibrium results for detection thresholds under operational gating.

\subsection{Graph and Taxonomy Structure}

\begin{figure}[ht]
    \centering
    \includegraphics[width=0.75\linewidth]{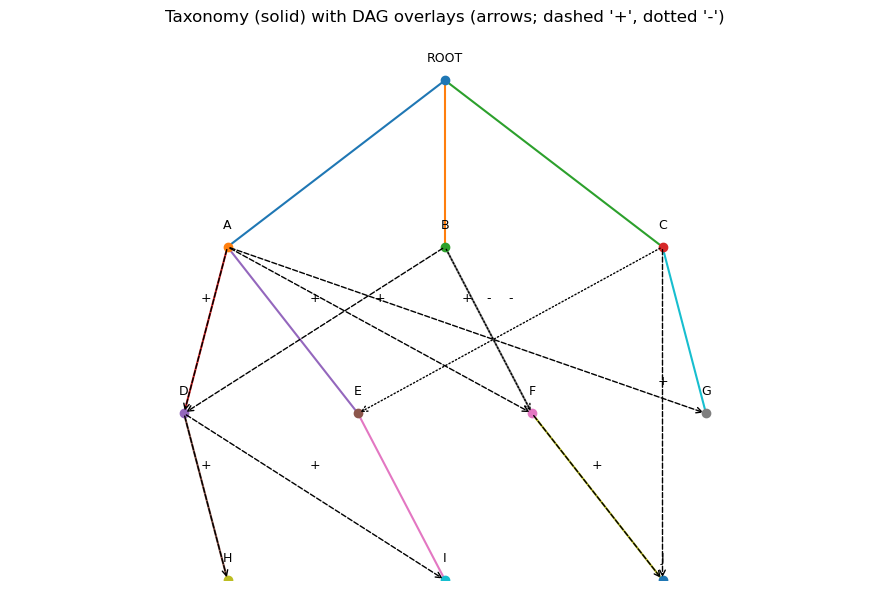}
    \caption{The taxonomy structure for the simulation study, of depth 4 leading off from a root node at the top. There are 12 total nodes in the tree, indexed by letters A-J and the ROOT node.}
    \label{fig:simulation_taxonomy_dag}
\end{figure}

We represent the detection network as a \emph{directed acyclic graph} (DAG) whose nodes correspond to detectors and whose edges reflect \emph{operational conditioning} relationships. Parent--child relationships are instantiated according to a given taxonomy of cyber incidents, and each child's operational firing condition is determined by a gate type:
\begin{itemize}
    \item \textbf{AND gate}: detector fires only if its internal rule is positive \emph{and} all parents in the \textbf{AND} set fire.
    \item \textbf{NOT gate}: detector fires only if its internal rule is positive \emph{and} all parents in the \textbf{NOT} set do not fire.
\end{itemize}
The \textbf{AND} and \textbf{NOT} sets are indicated by an indicator taking values of "+" and "-", respectively, in the code. An illustration of the DAG for the simulations is provided in Figure \ref{fig:simulation_taxonomy_dag}; with $10$ edges corresponding to $8$ \textbf{AND} gates and $2$ \textbf{NOT} gates.

These gate semantics are enforced in the code base using the DAG utilities, outlined in \texttt{PerformanceGraphsClasss.py} in the code-base in Section \ref{sec5}, with cycle prevention and topological ordering. We don't model the gates directly as binary, but model them as \emph{influence scores} from $[0,1]$, where values on the boundary indicate hard rules. For the simulations the influence scores are set to $.2$ for both \textbf{AND} and \textbf{NOT} gates across all nodes. We also introduce a strength parameter indicating the interconnectedness of all edges in the network $\gamma>0$, which in our simulations is generated from a $2*\textrm{Beta}(1,1)$ for all edges in the graph, seeded appropriately. This indicates high interdependence.

Specifically, $\gamma$ is incorporating into the neighbor calculations for utility as an additive term in the utility function weighting the relationship of neighbors in the Moral Graph of the DAG (\cite{koller2009probabilistic}.

\subsection{Generative Model for Detector Scores}

Each detector is constructed out of a \texttt{TaxonomyNode} class object outlined in \texttt{TaxonomyNodeClass..py}, in the same linked code-base, and is endowed with a latent score distribution for positive and negative cases. In the current code base, these scores are indicated by a prevalence parameter between $[0,1]$. This control the type $\textrm{1}$ and $\textrm{2}$ error for each $\lambda$ value for the associated node. The \emph{base rate} (prevalence of positives) is fixed per detector. Precision and recall are computed to be functions defined as a function of the base sensitivity and specificity at $\lambda$ (which in this scenario comes from:

\begin{verbatim}
def clamp(x: float, lo: float = 0.0, hi: float = 1.0) -> float:
    return max(lo, min(hi, x))

def base_sensitivity(self, lam: float) -> float:
    return math.sqrt(clamp(1.0 - lam))

def base_specificity(self, lam: float) -> float:
    return math.sqrt(clamp(lam))
\end{verbatim}

We then incpororate the DAG component, and a child in the DAG can be influenced by it's parents in the DAG as:
\begin{verbatim}
# --- Effective sensitivity/specificity with parental influence ---
    def effective_s_and_t(self, lam: float, neighbor_s: Dict[str, float], neighbor_t: Dict[str, float]) -> Tuple[float, float]:
        s0 = self.base_sensitivity(lam)
        t0 = self.base_specificity(lam)

        if not self.in_edges:
            return s0, t0

        pos_par_s = [neighbor_s[nm] for nm, sgn in self.in_edges.items()\
                      if sgn > 0 and nm in neighbor_s]
        pos_par_t = [neighbor_t[nm] for nm, sgn in self.in_edges.items()\
                      if sgn > 0 and nm in neighbor_t]
        neg_par_s = [neighbor_s[nm] for nm, sgn in self.in_edges.items()\
                      if sgn < 0 and nm in neighbor_s]
        neg_par_t = [neighbor_t[nm] for nm, sgn in self.in_edges.items()\
                      if sgn < 0 and nm in neighbor_t]

        # Means centered around 0.5 to create monotone shifts consistent with Theorem 2 / Prop 1
        def centered_mean(xs):
            if not xs: return 0.0
            return sum(x - 0.5 for x in xs) / len(xs)

        s = s0 + self.pos_weight * centered_mean(pos_par_s) -\
              self.neg_weight * centered_mean(neg_par_s)
        t = t0 + self.pos_weight * centered_mean(pos_par_t) -\
              self.neg_weight * centered_mean(neg_par_t)

        return clamp(s), clamp(t)
\end{verbatim}

The baseline sensitivity and specificity of a node are modified based on the ancestors/children in the Taxonomy.

This setup makes it possible to control:
\begin{itemize}
    \item \textbf{Signal strength} between positives and negatives (affecting achievable precision/recall).
    \item \textbf{Interconnectedness} in the DAG (how much a parent's firing rate impacts a child's base rate through gating).
    \item \textbf{Noise} in parent--child taxonomy relationships (for stochastic-taxonomy variants).
\end{itemize}

\subsection{Threshold Optimization Regimes}
\label{subsec:optimization_regimes}
The simulations compare three optimization regimes for threshold selection:
\begin{enumerate}
    \item \textbf{Single-pass staged optimization}: Thresholds are optimized in topological order of the DAG, with each node's threshold chosen to maximize its own utility
    \[
    U_D = a_D \cdot \mathrm{Precision}_D + b_D \cdot \mathrm{Recall}_D
    \]
    given its parents' fixed thresholds.

    \item \textbf{Forward--backward sweep until convergence}: A local optimization procedure sweeps forward and backward through the DAG repeatedly, updating each node's threshold to its current best response until the threshold vector converges within a small tolerance.

    \item \textbf{Global (equilibrium) search}: A coarse grid search over all thresholds in the action sets, selecting the profile that maximizes the joint objective (or meets the equilibrium conditions). In practice, this is tractable only for small networks due to combinatorial growth.
\end{enumerate}

These methods correspond directly to the theoretical constructs: (1) single-pass staged optimization models a one-shot myopic strategy; (2) sweeps approximate best-response dynamics; (3) the global search stands in for the pure-strategy Nash equilibrium guaranteed by the existence theorem.

\subsection{Evaluation Metrics}

The primary outcome is each node's utility value under the three regimes, plotted side-by-side in \textbf{boxplots} across repeated random instantiations of the generative model. The simulations also record precision and recall separately to verify the monotonicity and continuity predictions:
\begin{itemize}
    \item \textbf{Continuity}: sweeping thresholds produces smooth changes in measured metrics, consistent with Theorem~1.
    \item \textbf{Monotonicity}: for AND gates, increasing parent precision tends to increase child precision (and decrease recall), matching Theorem~2 and Proposition~1.
    \item \textbf{Hierarchy effect}: when gating respects the taxonomy structure, children under their parents show higher precision and lower recall in the deterministic-taxonomy limit (Theorem~3).
\end{itemize}

\subsection{Repetition and Randomization}

For each scenario, the simulation is repeated multiple times (with independent random seeds for score generation) to capture variability due to stochastic sampling. Boxplots aggregate these replicates, showing the distribution of utility improvements from local methods toward equilibrium.

\subsection{Implementation Notes}

\begin{itemize}
    \item All DAG operations, gating, and staged/sweep solvers are implemented in \texttt{PerformanceGraphsClass.py}.
    \item Node attributes, including base rates and score distributions, are defined in \texttt{TaxonomyNodeClass.py}.
    \item Visualization code (boxplots, convergence traces) is contained in the main script or notebook.
    \item Parameters such as base rate, signal strength, and noise can be varied to explore sensitivity, though in the current code these are fixed in the main demonstration.
\end{itemize}

\subsection{Nash Equilibrium Improvements}
\label{subsec4b}

\begin{figure}[ht]
    \centering
    \includegraphics[width=0.95\linewidth]{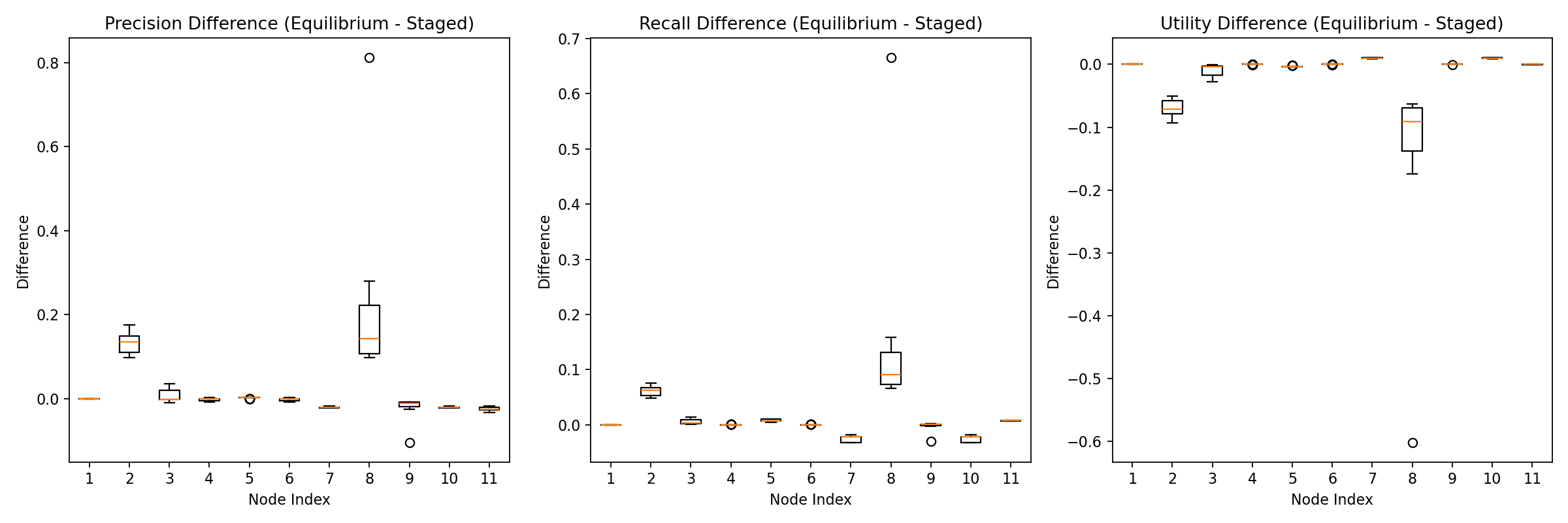}
    \caption{Boxplots showing the difference in Nash equilibrium and staged single pass frozen optimization. Precision, Recall, and Utility are shown for for $a=.4$ and $b=.6$. Values generated across 20 replications and varying values for the interconnectedness parameter in our model between [0,2] randomly generated uniformly.}
    \label{fig:diff_boxplot}
\end{figure}

\begin{figure}[ht]
    \centering
    \includegraphics[width=0.95\linewidth]{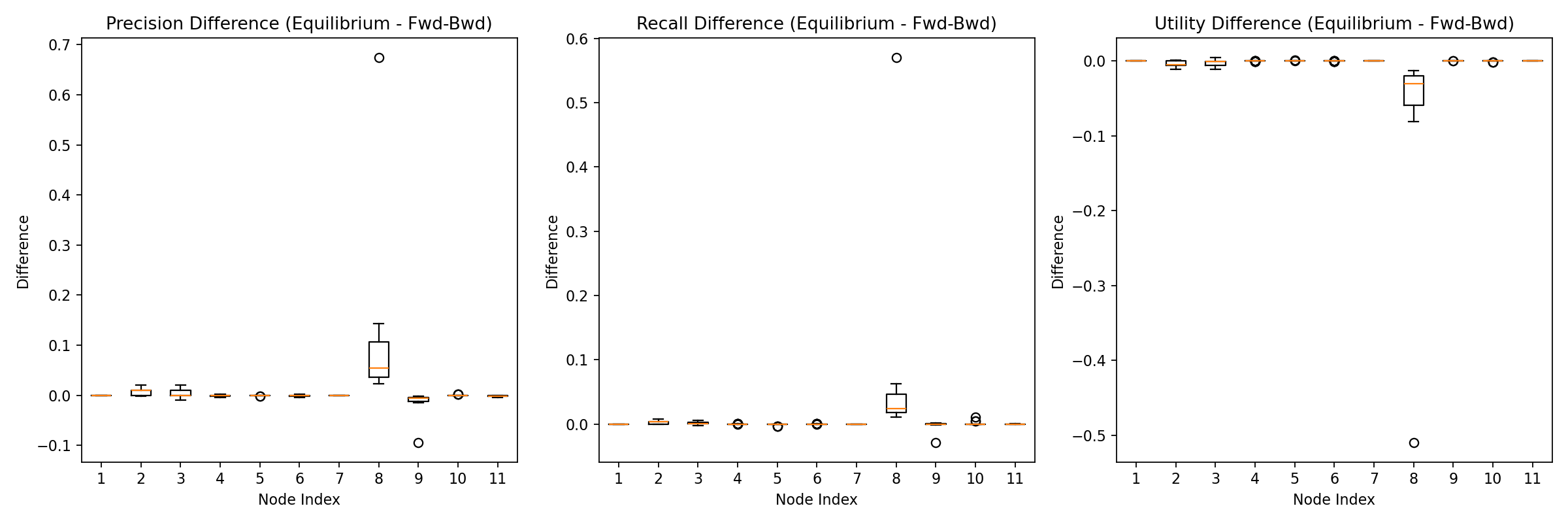}
    \caption{Boxplots showing the difference in Nash equilibrium and staged forward backward non-frozen optimization. Precision, Recall, and Utility are shown for for $a=.4$ and $b=.6$. Values generated across 20 replications and varying values for the interconnectedness parameter in our model between [0,2] randomly generated uniformly.}
    \label{fig:forward_backward}
\end{figure}

To illustrate the Nash equilibrium result in a simple scenario, assume that all paths downward from this tree at any depth (including only the root node) are local detection subsets indicating an attack strategy. This is in line with the outline previously presented; in the sense that modifying any particular attack strategy for optimal performance might affect the other attack strategy detections. We illustrate the three scenarios outlined in Section \ref{subsec:optimization_regimes}, first comparing regime (1) to (3), then comparing (2) to (3). The results are outlined in Figure \ref{fig:diff_boxplot} for comparison between single pass frozen optimization and the nash equilibrium, and in Figure \ref{fig:forward_backward} for the iterated forward-backward non-frozen pass. In our setting the criteria for convergence of the nonfrozen solver was maximum iteration of $1000$ and a maximum $\Delta\lambda$ of $10^{-6}$.

We can see that the equilibrium improves performance in some nodes dramatically, at a small trade off with other nodes for the single pass in Figure \ref{fig:diff_boxplot}. Figure \ref{fig:forward_backward} illustrates another interesting dynamic. It appears that as we run he forward-backward pass over multiple iterations it slows diminishes the difference between the equilibrium and the single pass, converging to the Nash equilibrium. This is consistent with broader dynamic system theory which shows that under all actors behaving individually in their best interests the system converges to the Nash equilibrium given enough time \cite{dynamic_system}.

\subsection{Illustration of the Results Through Simulations}
\label{subsec:results_simulations}

\begin{figure}[H]
    \centering
    \includegraphics[width=0.66\linewidth]{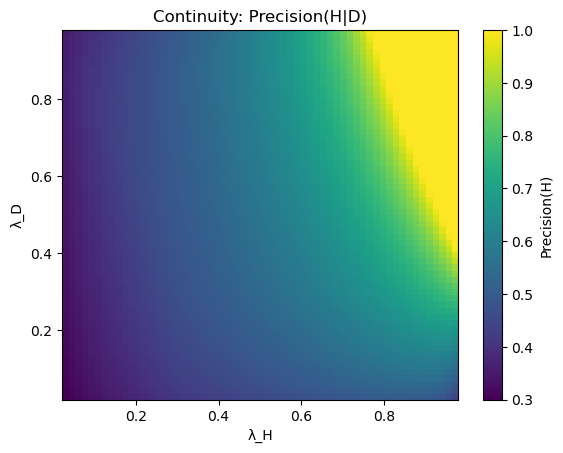}
    \caption{A figure illustrating the smoothness of a synthetic detection with respect to thresholding parameters}
    \label{fig:continuity}
\end{figure}

We proceed to illustrate the other theorems with simulations. Figure \ref{fig:continuity} illustrates the smoothness of the precision surface with respect to the thresholding parameters. We can see that $Precision(H|D)(\lambda_H,\lambda_D)$ is continuous both in $\lambda_H$ and $\lambda_D$. Additionally, this theorem illustrates Theorem \ref{thm:mono-and}. As $\lambda_D\rightarrow 1$ $\textrm{Precision}(D)(\lambda_D)$ increases, and so does $\textrm{Precision}(H|D)(\lambda_H,\lambda_D)$ across all values of $\lambda_H$. Heading in the opposite direction, as $\lambda_D \rightarrow 0$, we get the result for recall in Theorem \ref{thm:mono-and} as well. The inverse relationship between precision and recall gives us the converse result for recall outlined in the same theorem.

\begin{figure}[H]
    \centering
    \includegraphics[width=0.66\linewidth]{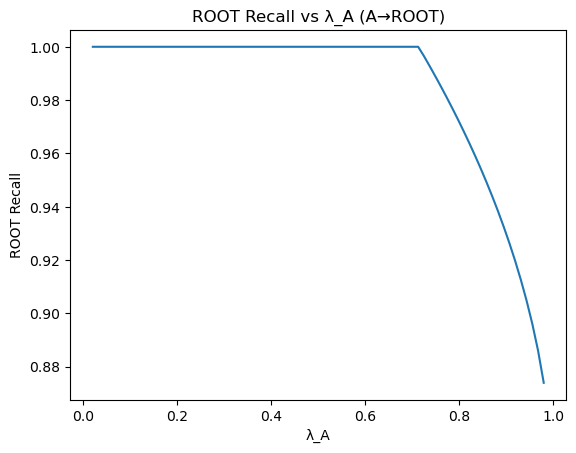}
    \caption{A figure illustrating Theorem \ref{thm:hier-det}. As the Precision of the child in the taxonomy (which the parent in the taxonomy is conditioned upon) increases, the recall of the parent in the taxonomy (which is the child in the DAG) goes down monotonically. The inverse is true in the opposite direction.}
    \label{fig:child_conditioning}
\end{figure}

Figure \ref{fig:child_conditioning} illustrates the deterministic taxonomy descendant conditioning outlined in Theorem \ref{thm:hier-det}.

\section{Discussion}\label{sec:discussion}
This work has introduced a framework for integrating cyber incident taxonomies into networks of detections in order to optimize their joint performance under precision–recall trade-offs. By viewing each detection as a noncooperative agent with its own performance objectives, we linked the tuning of detection thresholds to game-theoretic equilibrium analysis.

Our key theoretical contributions are:
\begin{itemize}
\item \textit{Smoothness and Monotonicity under gating}
We formally established conditions under which precision and recall vary continuously with detection thresholds, and how gating relationships (e.g., AND, NOT) induce monotonic effects across parent–child detection pairs.

\item \textit{Taxonomy-Structured Performance Bounds}
For deterministic taxonomies, conditioning a parent in the taxonomy on a descentant in the taxonomy results in changes in precision or recall; depending on the gating. We generalized this to stochastic taxonomies, quantifying robustness to noise.

\textit{Existence of Nash Equilibrium in Threshold Selection}
We proved that under mild continuity and quasi-concavity conditions, there exists a pure-strategy Nash equilibrium in the threshold configuration space. This equilibrium represents a stable configuration where no detector can unilaterally improve its utility.

From a practical perspective, the equilibrium result suggests that:
Real-world detection networks can be tuned iteratively to approach optimal joint performance without requiring central coordination.
In simulation, forward–backward non-frozen optimization approximates equilibrium performance over time, aligning with convergence results from dynamic systems theory

\item \textit{The simulation results illustrate that:}
\begin{enumerate}
\item Single-pass local optimization underperforms compared to the global equilibrium.
\item Iterative forward–backward optimization converges toward the equilibrium, validating the theoretical predictions.
\item The shape of the precision–recall trade-off surfaces can guide practitioners in selecting threshold adjustments most likely to yield performance gains.
\end{enumerate}
\item \textit{Future Directions}
\begin{enumerate}
\item Extending the framework to multi-class taxonomies and multi-objective utility functions.
\item Incorporating cost-sensitive metrics (e.g., weighted false positives/negatives).
\item Exploring learning-based approaches for equilibrium estimation in large-scale detection graphs.
\end{enumerate}
\end{itemize}
We believe that, by unifying taxonomy-based detection design with game-theoretic optimization, this work opens the door for more adaptive and resilient cyber defense systems that operate efficiently even in interconnected and dynamic environments.

\section*{Supplementary information}
\label{sec5}
Code is available at the following github repository: \url{https://github.com/rswarnick1/Performance_Graphs}.

\section{Acknowledgements}

We would like to thank the Microsoft Security Research Leadership Team for permission to publish. Code was generated with the assistance of GPT5.

\begin{appendices}

\section{Appendix}\label{appendix}

\subsection{Proof of Theorem \ref{theorempartialcontinuity}}
\begin{statement}
[\textbf{Continuity/semi-continuity under gating}]
Under (A1)–(A3) and fixed $\lambda_{-D}$, the map $\lambda_D \mapsto \mathrm{Rec}_{D^{\mathrm{op}}}(\lambda_D\,;\,\lambda_{-D})$ is continuous on $\Lambda_D$. 
Moreover, $\lambda_D \mapsto \mathrm{Prec}_{D^{\mathrm{op}}}(\lambda_D\,;\,\lambda_{-D})$ is upper semicontinuous on $\Lambda_D$, and continuous wherever $\mathbb{P}(D^{\mathrm{op}}_{\lambda}=P)$ stays bounded away from $0$.
\end{statement}
\label{proofpartialcontinuity}
\begin{proof}
Both recall and the numerator of precision are expectations of bounded indicator functions whose arguments depend continuously on $\lambda_D$; the denominator inherits upper semicontinuity from dominated convergence. The ratio of a continuous (respectively upper semicontinuous) numerator and a denominator bounded away from $0$ is continuous (respectively upper semicontinuous); on the constraint set where the denominator is positive, upper semicontinuity follows by standard closure arguments.

$\blacksquare$.
\end{proof}
\subsection{Proof of Theorem \ref{thm:mono-and} and Proposition \ref{prop:mono-not}}

\begin{statement}
[\textbf{Monotonicity of precision/recall under \textsc{AND} gating}]
Fix a detector $D$ with parents $\mathcal{P}(D)$ under \textsc{AND} gating. Suppose that for every parent $C\in\mathcal{P}(D)$ we move $\lambda_C$ to weakly increase $\mathrm{Prec}_C$ (with recall not increasing). Then for fixed $\lambda_D$, $\mathrm{Prec}_{D^{\mathrm{op}}}$ weakly increases and $\mathrm{Rec}_{D^{\mathrm{op}}}$ weakly decreases. 
\end{statement}

\begin{statement}
[\textbf{Monotonicity under \textsc{NOT} gating}]
Under \textsc{NOT} gating, if $\mathrm{Prec}_C$ weakly increases (with recall not increasing) for each parent $C$, then $\mathrm{Prec}_{D^{\mathrm{op}}}$ weakly \emph{decreases} and $\mathrm{Rec}_{D^{\mathrm{op}}}$ weakly \emph{increases}.
\end{statement}
\label{proofmono}
\begin{proof}
Treat the operational event as set intersections/unions and use inclusion relations of conditioning events; apply isotonicity of conditional probabilities (\cite{angeli2003monotone}) with respect to set containment, keeping $\lambda_D$ fixed.
$\blacksquare$
\end{proof}

\subsection{Proof of Theorem \ref{thm:hier-det} and Theorem \ref{thm:hier-stoch}}\label{proofroot}
\begin{statement}
[\textbf{Deterministic taxonomy idealization}]
Assume a lossless taxonomy such that every child label implies its ancestor: $Y_n=P \Rightarrow Y_m=P$ for all ancestors $m \prec n$ (no noise). Then for any child $D_n$ gated by parent $D_m$ via \textsc{AND},
\[
\mathrm{Prec}_{C_n^{\mathrm{op}}} \ge \mathrm{Prec}_{D_m}, \qquad 
\mathrm{Rec}_{C_n^{\mathrm{op}}} \le \mathrm{Rec}_{D_M}.
\]
\end{statement}

\begin{statement}
[\textbf{Stochastic taxonomy, robustness}]
If $\mathbb{P}(Y_m=P \mid C_n=P) \ge 1-\epsilon$ for all thresholds in the action sets, then
\[
\mathrm{Precision}_{C_n^{\mathrm{op}}} \ge \mathrm{Precision}_{C_m} - O(\epsilon), \qquad 
\mathrm{Recall}_{C_n^{\mathrm{op}}} \le \mathrm{Recall}_{C_m} + O(\epsilon).
\]
\end{statement}

We use the subscript $n$ for thresholds, detections, and signals to indicate $\epsilon_1,\dots,\epsilon_n$, and similarly $m$ for $\epsilon_1,\dots,\epsilon_m$.

Note that the event of a $Y_n = P$ in $T_n$ implies an event of a $P$ for $Y_m$ in $T_m$ with probability $1$ . 

Note from equation \ref{limit_precision} that:
\begin{equation}
Precision(D_n) = \lim_{N\rightarrow \infty}\frac{TP(D_n)(\lambda_{n})}{TP(D_n)(\lambda_n) + FP(D_n)(\lambda_n)}
\end{equation}

we then have that:

\begin{align}
Precision(D_n)(\lambda_n) &> \frac{P(D_n=P,Y_n=P | Y_m = P, D_m = P)(\lambda_n, \lambda_m)}{P(D_n = P| D_m = P, Y_m = P)(\lambda_n,\lambda_m)}\\
Precision(D_n)(\lambda_n) &>P(Y_n=P| Y_m = P, D_m = P, D_n = P)(\lambda_n, \lambda_m) \label{eq:proofroot}
\end{align}

But note that $_m = P \Rightarrow Y_n=P$ with probability $1-\epsilon$ logically, as $T_n$ stochastically contains $T_m$ with probability $1-\epsilon$. This means that the probably on the right hand side of inequality \ref{eq:proofroot} is at least $1-\epsilon$. Using the law of total probability gives us $Precision(D_n)($

This gives us the correct lower bound for Theorem  \ref{thm:hier-stoch}. Taking the limit as $\epsilon\rightarrow 0$ gives us the idealized case for Theorem\ref{thm:hier-det}.

\subsection{Proof of Theorem \ref{thm:existence}}\label{proofequilibria}
\begin{statement}
[\textbf{Existence of pure-strategy equilibrium}]
Let $\{\Lambda_D\}_{D=1}^N$ be nonempty, compact, convex intervals. For each $D$, define the utility
\[
U_D(\lambda_D;\lambda_{-D}) = a_D \,\mathrm{Prec}_{D^{\mathrm{op}}}(\lambda_D;\lambda_{-D}) 
+ b_D \,\mathrm{Rec}_{D^{\mathrm{op}}}(\lambda_D;\lambda_{-D}),
\quad a_D,b_D \ge 0,\; (a_D,b_D)\neq (0,0).
\]
Assume (A1)–(A3) and that for every fixed $\lambda_{-D}$, $U_D(\cdot;\lambda_{-D})$ is quasi-concave on $\Lambda_D$ and $U_D$ is continuous on $\prod_D \Lambda_D$.

Then this framework is a noncooperative game with an equilibrium point \cite{debreu1952social,glicksberg1952further,fan1952minimax} where appropriate thresholds can be selected where no detection can be improved without other detections wanting to change their thresholds for improvement. 
\end{statement}
\begin{proof}
We show that the detection threshold optimization problem is a noncooperative game satisfying the hypotheses of the Glicksberg extension of the Kakutani fixed-point theorem.

\begin{enumerate}
\item 
\begin{itemize}\textit{Players and Strategy Sets}
\item Each Detector $D$ is a player.
\item The strategy set for $D$ is $\Lambda_D$, which is nonempty, compact, and convex by assumption.
\end{itemize}
\item
 \begin{itemize}\textit{Utility Functions}
\item By (A1)-(A3), $\textrm{Precision}_D^{op}$ is upper semi-continuous, and $\textrm{Recall}_D^{op}$ is continuous on $\lambda_D$ for fixed $\lambda_{-D}$
\item Therefore $U_D(\circ;\lambda_{-D)}$ is continuous and quasi concave on $\Lambda_D$.
\end{itemize}
\item 
\begin{itemize}\textit{Best Response Correspondence}
\item For each fixed $\lambda_{-D}$, define the best response set:
\begin{equation}
BR_D(\lambda_{-D}) = \argmax_{\lambda_D \in \Lambda_D} U_D(\lambda_D;\lambda_{-D})
\end{equation}
\item Quasi-concavity of $U_D$ ensure that $BR_D(\lambda_{-D})$ is convex valued and nonempty.
\item Continuity ensures that $BR_D$ is upper hemicontinuous.
\end{itemize}
\item
 \begin{itemize}\textit{Product Correspondence}
\item Define $BR(\lambda) = \prod_{D\in \mathcal {D}}BR_D(\lambda_{-D})$
\item $BR$ maps $\Lambda$ into itself with nonempty, convex values and is upper hemicontinuous.
\end{itemize}
\item 
\begin{itemize}\textit{Existence of Equilibrium}
\item By Kakutani's fixed point theorem, there exists $\lambda^*\in\Lambda$ such that:
\begin{equation}
\lambda^*\in BR(\lambda^*)
\end{equation}
This $\lambda^*$ is a pure strategy Nash equilibrium. \cite{debreu1952social,glicksberg1952further,fan1952minimax}.
\end{itemize}
\end{enumerate}
$\blacksquare$
\end{proof}




\end{appendices}
\bibliographystyle{unsrt}  
\bibliography{Detection_Taxonomy_Project}  

\end{document}